\documentstyle[aps,preprint]{revtex}
\draft

\begin{document}

\title{Thermodynamic entropy and excess information  loss in 
dynamical systems with time-dependent Hamiltonian}

\author{Shin-ichi Sasa and Teruhisa S. Komatsu}

\address{
Department of Pure and Applied Sciences, 
          University of Tokyo, \\
         Komaba, Meguro-ku, Tokyo 153, Japan}

\date{July 6, 1998}

\maketitle

\begin{abstract}
We study a dynamical system with  time dependent Hamiltonian by numerical 
experiments so as to find a relation between thermodynamics and chaotic 
nature of the system. Excess information loss, defined newly based on 
Lyapunov analysis, is related to the increment of thermodynamic 
entropy. Our numerical results suggest that the positivity of  entropy 
increment is expressed by the principle of the minimum excess information 
loss. 
\end{abstract}



\vskip5mm



Boltzmann found a famous formula which relates thermodynamic 
entropy with a number of states in microscopic systems.
While the formula was practically well-established, 
there have been discussions on its justification over the century
\cite{Lebo}.
In particular,  understanding  the origin of irreversibility 
seems to be controversial even to the present knowledge.
On the other hand, owing to development of computational circumstance,
we can easily investigate numerical solutions to equations of motion
in Hamiltonian systems with a high degrees of freedom. 
Thus, through  numerical experiments, it may be possible to 
get new insights for the relation 
between thermodynamics and dynamical systems. In particular, we are
concerned with a question how thermodynamic entropy is related to
notions of dynamical system theory.


We  study  numerically  a Fermi-Pasta-Uram (FPU)  model
\cite{FPU},
whose Hamiltonian is given by
\begin{equation}
H(\{q _i\}, \{p_i \};g)= 
\sum_{i=1}^N [{1 \over 2} p_i^2+
{1\over 2}(q_{i+1}-q_{i})^2+{g \over 4}(q_{i+1}-q_{i})^4],
\label{hamil}
\end{equation}
where $g$ is a time dependent parameter. Then, 
the evolution equations for  $(\{q_i\},\{p_i\})$ are written as
\begin{eqnarray}
{d q_i \over dt}  &=& p_i, 
\label{evol:q} \\
{d p_i \over dt}  &=& (q_{i+1}-q_{i})+g(q_{i+1}-q_{i})^3
-(q_{i}-q_{i-1})-g(q_{i}-q_{i-1})^3.
\label{evol:v}
\end{eqnarray}
We assume periodic boundary conditions, that is, $q_0$ and $q_{N+1}$ 
in Eq.(\ref{evol:v}) satisfy $q_0 = q_{N}$ and $q_{N+1}= q_{1}$.
We also assume that the values of conserved quantities 
$\sum_i q_i$ and $\sum_i p_i$ are zero. We solve numerically
Eqs.(\ref{evol:q}) and (\ref{evol:v}) by the 4-th order symplectic 
integrator method with a time step $\delta t=0.005$ \cite{symp}.
When we are concerned with the system in the thermodynamic limit,
numerical calculation should be done with several values of $N$.
In the argument below, results for $N=5$ and $N=20$ will be presented.


We first discuss the case that $g$ takes a constant value, say $g_0$.
We assume that initial conditions are chosen from the
micro-canonical ensemble with an energy $E_0$.
When $ E_0g_0$ is sufficiently large, the system exhibits 
high-dimensional chaos.  As an example of such parameter values,
$(E_0,g_0)=(1.0, 10.0)$ is assumed.
The nature of high-dimensional chaos is characterized by
Lyapunov spectrum and Kolmogorov-Sinai (KS) entropy. 
We first review briefly 
how to obtain them numerically. This is a well-known fact, but  
will help us to introduce a new quantity ``excess information loss'' later.


Let a solution to Eqs.(\ref{evol:q}) and (\ref{evol:v}) denote
$
\Gamma(t) = (q_1(t),p_1(t),q_2(t),p_2(t),\cdots,q_N(t),p_N(t)).
$
In Lyapunov analysis, divergence and convergence of distance 
between neighboring orbits are described by time evolution  
of a set of  $2N$ orthogonal unit vectors, $\{ {\bf e} _i\}_{i=1}^{2N}$,
each of which obeys the linearized equations of 
Eqs.(\ref{evol:q}) and (\ref{evol:v}). 
By employing the Gram-Schmidt procedure, which was developed as 
a numerical calculation method for Lyapunov exponents \cite{SN},
the evolution of  ${\bf e} _i$, $ \hat T(t,0) {\bf e} _i$, is expressed by
\begin{equation}
\hat T(t,0) {\bf e} _i =\sum_{j} \hat F(t,0){\bf e} _j R_{ji}(t,0),
\label{evol:e}
\end{equation}
where $R_{ij}$ is the $(i,j)$ element of an upper triangle
matrix $\hat R$, and $\hat F(t,0)$ is an orthogonal matrix. 
We can find the matrices $\hat R$ and $\hat F$ so that
the diagonal elements of $\hat R$  are positive.
Note that 
$\{ \hat F(t,0) {\bf e} _i \}_{i=1}^{2N} $ gives an orthogonal set
defined in the tangent space at the point $\Gamma(t)$.
{}From the definition of the matrix $\hat R$, 
the $j$-dimensional parallelepiped volume spanned 
by $\{\hat T(t,0) {\bf e} _i \}_{i=1}^j$ is calculated as
\begin{equation}
\Omega_j(t,0;\{{\bf e} _i\})= \Pi_{i <j}  R_{ii}(t,0).
\end{equation}
The $i$-th Lyapunov exponent $\lambda_i$ is then defined  by
\begin{equation}
\lambda_i = \lim_{t\rightarrow \infty}{1\over t}
\log {\Omega_i(t,0;\{{\bf e} _i\})\over \Omega_{i-1}(t,0;\{{\bf e} _i\})}.
\end{equation}
All Lyapunov exponents for the model in question
are shown in Fig. \ref{fig1}.  Note that  there are 
four zeros due to the periodic boundary conditions.
Further, the KS entropy $h$ is given by \cite{Pesin}
\begin{equation}
h=\sum_{\lambda_i >0} \lambda_i.
\end{equation}
Since the number of positive Lyapunov exponents is $N-2$,
the KS entropy is expressed by
\begin{equation}
h=\lim_{t\rightarrow \infty}
{1\over t}\log\Omega_{N-2} (t,0;\{{\bf e} _i\}).
\label{KS:def}
\end{equation}
We also  define  a special orthonormal set  
$\{ {\bf e} _i^* \}_{i=1}^{2N} $  at a point $\Gamma(0)$ by 
\begin{equation}
{\bf e} _i^*= \lim_{t_b \rightarrow -\infty} \hat F(0,t_b){\bf e} _i,
\label{lv:def}
\end{equation}
where $\{ {\bf e} _i \} $ is an orthonormal set 
defined in the tangent space at $\Gamma(t_b)$. 
${\bf e} _i^*$ corresponds to the $i$-th Lyapunov vector at a point 
$\Gamma(0)$.


Now, we consider the case that the value of $g$ is changed in time 
from $g_0$ to $g_1$.
Especially, we pay attention to two limiting cases of the parameter 
change, quasi-static process and instantaneous switching process.
In a quasi-static limit, the parameter is changed in a slower way
than  all physical time scales. 
Since thermodynamic entropy in isolated systems should be kept at constant
in quasi-static processes,  equi-entropy lines in the $(E,g)$ plane 
can be obtained  numerically by quasi-static processes.
Figure \ref{fig2} shows the equi-entropy line  through $(E_0,g_0)$,
which is denoted by $E=E_{qs}(g)$. We also define $E_*(E,g)$ 
by a requirement that $(E_*,0)$ is on  the equi-entropy line 
through $(E,g)$. Since
the FPU model with $g=0$ is nothing but the harmonic oscillator model,
we can evaluate the entropy at $(E,g)$ as 
\begin{equation}
S(E,g)=S(E_*(E,g),0)=(N-2)\log E_*(E,g),
\label{ent-bolt}
\end{equation}
where  an additive constant with respect to $E_*$ is omitted, and
the Boltzmann constant is assumed to be unity.

Instantaneous switching is the other extreme limit of the parameter change.
The value of $g$ is  changed  instantaneously from $g_0$ to $g_1
=g_0+\Delta g$  at $t=0$. Then, the energy after the switching
becomes $E_1$, whose value depends on the choice of initial conditions.
The entropy difference $\Delta S$ can be defined by
\begin{equation}
\Delta S=S(E_1,g_1)-S(E_0,g_0),
\end{equation}
where $ S $ is  calculated by the formula Eq.(\ref{ent-bolt}).
In Fig. \ref{fig3}, the average of the entropy difference
over initial conditions $ \langle \Delta S \rangle$ was plotted  
against $\Delta g$. 
In the thermodynamic limit, the relative fluctuation of $E_1$ becomes
negligible and then $E_1$ can be identified with the averaged value 
$\langle E_1 \rangle $.  
Note  that  $\langle E_1 \rangle$ satisfies 
\begin{equation}
\langle E_1 \rangle -E_0=
\left( d E_{qs}(g) \over d g \right)
(g_1-g_0).
\label{jump}
\end{equation}
That is, $\langle E_1 \rangle $ is determined in such a way that 
$(\langle E_1 \rangle ,g_1)$ is in the straight line contacting to 
the equi-entropy  line $E=E_{qs}(g)$ at $(E_0,g_0)$. 
(See Fig. \ref{fig2}.)  
We also confirmed this statement numerically. 
Since the equi-entropy line is convex in the $(E,g)$ plane,
the entropy difference between the states 
$(\langle E_1 \rangle,g_1)$ and $(E_0,g_0)$ turns out to be always positive.
In fact,  in Fig. \ref{fig3}, $S(\langle E_1 \rangle,g_1)-S(E_0,g_0)$ 
was shown against $\Delta g$. One may find that how $\Delta S$ approaches
to $S(\langle E_1 \rangle,g_1)-S(E_0,g_0)$ as $N$ is increased.


We now attempt to express $\Delta S$ at instantaneous switching
in terms of notions of dynamical system theory. 
One may naively expect that thermodynamic entropy 
is related to KS entropy. In fact, the extensivity of KS entropy suggests
a relation with thermodynamics. However, since KS entropy has the
same dimension as the inverse of time, the relation is not
straightforward. Further, we should notice that KS entropy 
measures the rate of information loss at a steady state in
dynamical systems, while thermodynamic entropy changes  only 
by external action. We thus expect that $\Delta S$ 
is related to  the change of information loss 
brought by the parameter switching. Motivated by this picture,
we next define the excess information loss.


Since the parameter is switched instantaneously at $t=0$,
the orbit is not smooth at this time.
We thus define a smooth orbits $\Gamma_{st}(t;g_1)$ in such a way
that $\Gamma_{st}(t;g_1)=\Gamma(t)$  for $t \ge 0$.
The values of $\Gamma_{st}$ in the other region are determined
by the requirement of the smoothness  of $\Gamma_{st}$ at $t=0$.
Note that the orbit $\Gamma_{st}(\ ;g_1)$ is 
on a new energy surface. Since the two orbits $\Gamma( )$ and 
$\Gamma_{st}(\ ;g_1)$ intersect at $\Gamma(0)$, two sets of Lyapunov vectors,
denoted by ${\bf e} _i^*(g_0)$ and ${\bf e} _i^*(g_1)$ respectively,
are defined there by the time evolution along the two orbits 
$\Gamma( )$ and $\Gamma_{st}(\ ;g_1)$ from a sufficiently long ago.
(See Eq.(\ref{lv:def}).)
Then, recalling the definition of KS entropy given by Eq.(\ref{KS:def}),
we can interpret $\log \Omega_{N-2} (t,0;\{{\bf e} _i^*(g_0)\})$ as
the actual information loss  during a time interval $[0,t]$, 
because the set of Lyapunov vectors $\{{\bf e} _i^*(g_0)\}$ at $t=0$ 
is associated with the actual orbit $\Gamma()$.  On the contrary, 
the quantity  $\log \Omega_{N-2} (t,0;\{{\bf e} _i^*(g_1)\})$ 
corresponds to  the information loss  for the virtual orbit 
$\Gamma_{st}(\ ;g_1)$ on the new energy surface.
We now define the 'excess information loss' brought by
the parameter change  as
\begin{equation}
H_{ex}=\lim_{t\rightarrow \infty}
[\log 
\Omega_{N-2} (t,0;\{{\bf e} _i^*(g_0)\})
-
\log
\Omega_{N-2} (t,0;\{{\bf e} _i^*(g_1)\})] .
\end{equation}


In Fig. \ref{fig4}, the average of $H_{ex}$ over initial conditions, 
$\langle H_{ex} \rangle $, was plotted against $\Delta g$.
Comparing this graph with that in Fig. \ref{fig3}, one may doubt that 
$\langle H_{ex} \rangle $  has  no direct relation with  
$\langle \Delta  S \rangle $.
In particular, behaviors for $\Delta g \rightarrow 0$ are
qualitatively  different: $\langle H_{ex}\rangle \sim O(\Delta g)$, 
while $\langle \Delta S \rangle \sim O((\Delta g)^2) $. 
This property implies that the excess information loss 
in the quasi-static process $g_0 \rightarrow g_1$, denoted by
$\langle H_{ex} \rangle_{qs}$, takes a nonzero value. 
$\langle H_{ex} \rangle_{qs}$ may  be calculated as
\begin{equation}
\langle H_{ex} \rangle_{qs}  =\int_{g_0}^{g_1} dg \Psi(E_{qs}(g),g),
\end{equation} 
where $\Psi$ is  determined by an infinitely small jump process
$g \rightarrow g+\delta g$ at the energy $E_{qs}(g)$.
Now, it seems 
natural to assume the decomposition of $\langle H_{ex} \rangle$
in the form
\begin{equation}
\langle H_{ex} \rangle = \langle H_{ex} \rangle_{qs}+ c
\langle \Delta S \rangle,
\label{hex-ds}
\end{equation}
where $c_0$ is a constant.  In order to check the validity of
Eq.(\ref{hex-ds}),  we performed the reversed experiments 
in which the parameter $g$ is changed from $g_1$ to $g_0$
with the initial energy $E_{qs}(g_1)$.  We then obtained 
$\langle H_{ex} \rangle' $ and $ \langle \Delta  S \rangle'$.
We assume the relation Eq. (\ref{hex-ds}) for the reversed experiment,
that is,
\begin{equation}
\langle H_{ex} \rangle' = \langle  H_{ex} \rangle_{qs}'+ c 
\langle \Delta S \rangle',
\label{hex-ds:rev}
\end{equation}
where
\begin{eqnarray}
\langle  H_{ex} \rangle_{qs}'  &=&
\int_{g_1}^{g_0} dg \Psi(E_{qs}(g),g), \\
&=& -\langle H_{ex} \rangle_{qs}
\label{minus}.
\end{eqnarray} 
Therefore, from Eqs.(\ref{hex-ds}), (\ref{hex-ds:rev}) and (\ref{minus}),
we obtain
\begin{equation}
\langle H_{ex} \rangle+\langle H_{ex} \rangle' =
c  ( \langle \Delta S \rangle + \langle \Delta S \rangle').
\label{hex-ds:sum}
\end{equation}
This relation was checked numerically. As shown in Fig. \ref{fig5},
Eq. (\ref{hex-ds:sum}) seems valid with $c=0.5$.
We thus expect
\begin{equation}
\langle H_{ex} \rangle = \langle H_{ex} \rangle_{qs}+ {1\over 2}
\langle \Delta S \rangle .
\label{final}
\end{equation}
This is the main claim of the present paper. It is worthwhile 
noting that the positivity of the entropy difference is expressed by
the principle of the minimum excess information loss.


In closing the paper, we introduce related studies.
Recently, a relation between phenomenology of nonequilibrium steady states 
and dynamical system theory has been discussed by employing 
deterministic models with a thermostat or with open boundaries
\cite{Gaspard,Evans,Gall,Ruelle}. In particular,
an entropy production ratio is expressed in terms of KS entropy.
This may have correspondence with Eq.(\ref{final}).
(Note that we never discuss nonequilibrium steady states, but
 transitions between equilibrium states.) 
Further, Oono and Palconi have developed  new phenomenology
of steady state thermodynamics\cite{Oono}, 
in  which 'excess heat' plays an important
role in the definition of the entropy at a nonequilibrium steady state. 
When we replace a term  'heat' in their theory 
by 'information',  we find much similarity between two theories.
In order to discuss relations among the apparently different  theories,
we need to present a mathematical proof of Eq.(\ref{final}) 
together with the positivity of $\Delta S$.


The authors express special  thanks to Y. Oono for thoughtful ideas on 
thermodynamics.  They also acknowledge K. Sekimoto, H. Tasaki, S. Takesue, 
M. Sano, K. Kaneko, T. Chawanya and N. Nakagawa for  discussions. 
This work was supported by grants from 
the Ministry of Education, Science, Sports and Culture of 
Japan, No. 09740305  and from National Science Foundation, 
No. NSF-DMR-93-14938.  One of the authors (T.S.K.) also  
acknowledges JSPS and RIKEN.




\begin{figure}
\caption{Lyapunov spectrum. $N=20$. 
$\Omega_j(t,0)$ at $t=10^5$ was measured to evaluate the Lyapunov exponents.
}
\label{fig1}
\end{figure}

\begin{figure}
\caption{Averaged energy after instantaneous switching processes
and the equi-entropy line through $(E,g)=(1.0,10.0)$.
}
\label{fig2}
\end{figure}

\begin{figure}
\caption{Entropy difference versus $\Delta g$. N=20. Solid line shows
$S(\langle E_1 \rangle, g_1)-S(E_0,g_0)$. Square and filled circle
symbols represent $\langle S( E_1 , g_1)\rangle -S(E_0,g_0)$ for $N=5$
and $N=20$, respectively. 
}
\label{fig3}
\end{figure}

\begin{figure}
\caption{$H_{ex}$ versus $\Delta g$. Square and filled circle
symbols represent the data for $N=5$ and $N=20$, respectively. 
$H_{ex}$ was evaluated with checking the convergence.
}
\label{fig4}
\end{figure}

\begin{figure}
\caption{$\langle H_{ex} \rangle+\langle H_{ex} \rangle'$
versus $ \langle \Delta S \rangle + \langle \Delta  S \rangle' $.
Square and filled circle symbols represent the data for $N=5$ and $N=20$, 
respectively. The dashed line shows  
$\langle H_{ex} \rangle+\langle H_{ex} \rangle'=1/2[
\langle \Delta S \rangle + \langle \Delta  S \rangle']. $
}
\label{fig5}
\end{figure}

\end{document}